\documentclass[12pt]{article}
\usepackage{amssymb}
\usepackage{graphicx}
\newcommand{\nc}{\newcommand}
\nc{\postscript}[2] 
{\setlength{\epsfxsize}{#2\hsize}\centerline{\epsfbox{#1}}}
\nc{\bg}{B. Grzadkowski}
\nc{\non}{\nonumber}
\nc{\hc}{\hbox {h.c.}} \nc{\re}{\hbox {Re}} 
\nc{\mev}{\hbox {MeV}} \nc{\gev}{\;\hbox {GeV}} \nc{\tev}{\;\hbox {TeV}}
\def\lsim{\mathrel{\raise.3ex\hbox{$<$\kern-.75em\lower1ex\hbox{$\sim$}}}}
\def\gsim{\mathrel{\raise.3ex\hbox{$>$\kern-.75em\lower1ex\hbox{$\sim$}}}}

\nc{\prd}[3]{{\it Phys.\ Rev.}\ {{\bf D{#1}} (#2), #3}}
\nc{\prl}[3]{{\it Phys.\ Rev.\ Lett.}\ {{\bf {#1}} (#2), #3}}
\nc{\plb}[3]{{\it Phys.\ Lett.}\ {{\bf B{#1}} (#2), #3}}
\nc{\npb}[3]{{\it Nucl.\ Phys.}\ {{\bf B{#1}} (#2), #3}}
\nc{\ptp}[3]{{\it Prog.\ Theor.\ Phys.}\ {{\bf {#1}} (#2), #3}}
\nc{\zfp}[3]{{\it Z.\ Phys.}\ {{\bf C{#1}} (#2), #3}}
\nc{\epj}[3]{{\it Eur.\ Phys.\ J.}\ {{\bf C{#1}} (#2), #3}}
\nc{\mpla}[3]{{\it Mod.\ Phys.\ Lett.}\ {{\bf A{#1}} (#2), #3}}
\nc{\rmp}[3]{{\it Rev.\ Mod.\ Phys.}\ {{\bf {#1}} (#2), #3}}
\nc{\ijmpa}[3]{{\it Int.\ J.\ of\ Mod.\ Phys.}\
               {{\bf A{#1}} (#2), #3}}
\nc{\etal}{{\it et al.}}
\nc{\Lsp}{\;\;\;\;\;\;\;\;\;\;}  \nc{\LLLsp}{\lspace \lspace}
\nc{\lsp}{\;\;\;\;\;\;}
\nc{\spac}{\;\;\;}
\nc{\noi}{\noindent}
\nc{\beq}{\begin{equation}}   \nc{\eeq}{\end{equation}}
\nc{\bea}{\begin{eqnarray}}   \nc{\eea}{\end{eqnarray}}
\nc{\baa}{\begin{array}}      \nc{\eaa}{\end{array}}
\nc{\bit}{\begin{itemize}}    \nc{\eit}{\end{itemize}}
\nc{\ben}{\begin{enumerate}}  \nc{\een}{\end{enumerate}}
\nc{\bce}{\begin{center}}     \nc{\ece}{\end{center}}

\def\lam{\lambda}

\def\eps{\epsilon}

\def\lcal{{\cal L}}
\def\sq2{\sqrt{2}}

\def\ph{\varphi}

\def\m4{m^4(\ph)}
\def\mn2{m_n^2}

\def\phic{\phi_c}
\def\v5{V^{(5)}}

\begin{document}

\vspace*{-1.8cm}
\noindent
\begin{flushright}
$\vcenter{
\hbox{{\footnotesize CERN-PH-TH/2004-003}}
\hbox{{\footnotesize IFT-02-04}}
\hbox{{\footnotesize UCD-04-14}}
}$
\end{flushright}

\vskip 0.4cm
\bce
{\large\bf Low-energy effective theory from \\a non-trivial scalar 
background in extra dimensions}
\ece

\vspace{0.2cm}
\bce
\renewcommand{\thefootnote}{\alph{footnote})}
{\sc Bohdan GRZADKOWSKI}\footnote{E-mail address:
\tt bohdan.grzadkowski@fuw.edu.pl}

\vspace*{0.1cm}
{\sl Institute of Theoretical Physics,  Warsaw University\\
Ho\.za 69, PL-00-681 Warsaw, Poland\\
and\\
CERN, Department of Physics,\\
Theory Division\\
1211 Geneva 23, Switzerland}

\vspace*{0.5cm}
{\sc Manuel TOHARIA}\footnote{E-mail address:
\tt mtoharia@lifshitz.ucdavis.edu}

\vspace*{0.1cm}
{\sl Department of Physics, University of California\\
Davis CA 95616-8677, USA}
\ece

\vspace*{0.5cm}
\centerline{ABSTRACT}
\vspace*{0.5cm}
Consequences of a non-trivial scalar field background 
for an effective 4D theory were studied in the context of one compact extra dimension.
The periodic background that appears within the (1+4)-dimensional
$\phi^4$ theory was found and the excitations
above the background  (and their spectrum) were determined analytically.
It was shown that the presence of the non-trivial solution leads to the existence of a 
minimal size of the extra dimension that is determined by the mass parameter of the scalar potential.
It was proved that imposing orbifold antisymmetry boundary conditions allows us to eliminate a negative 
mass squared Kaluza--Klein ground-state mode that otherwise would cause an instability of the system.
The localization of fermionic modes in the presence of the non-trivial background was discussed in great detail
varying the size of the extra dimension and the strength of the Yukawa coupling.
A simple exact solution for the zero-mode fermionic states was found and the solution for
non-zero modes in terms of trigonometric series was constructed. The fermionic mass spectrum, which reveals 
a very interesting structure, was found numerically. It was shown that the natural size of the extra dimension
is twice as large as the period of the scalar background solution.

\vspace*{0.2cm} 

\vfill

PACS:  11.10.Kk, 11.27.+d

Keywords:
anomalous extra dimensions, kink solutions, fermion localization\\

\newpage
\renewcommand{\thefootnote}{$\sharp$\arabic{footnote}}
\pagestyle{plain} \setcounter{footnote}{0}
\baselineskip=21.0pt plus 0.2pt minus 0.1pt

\section{Introduction}
\label{intro}

A typical habit, which has its roots in the (1+3)-dimensional 
Standard Model (SM) and its possible 4D alternatives,
is to assume that the Higgs boson (or any other scalar) vacuum expectation value (vev) has a constant value. 
In 4D theories, this is
a consequence of the requirement of the 4D translational invariance. However, 
in models of the electroweak interactions
built upon $1+(3+d)$-dimensional manifolds, the translation invariance is no longer required, as the
extra dimensions must be compact. Here we will consider the space-time manifold that is $M^{(3,1)}\times R$,
where $M^{(3,1)}$ is the usual Minkowski space, while $R$ is a compact manifold. 
The vev is no longer forced to be a constant by the (4D) symmetries that are typically imposed, and therefore 
it can have a non-trivial profile along the extra
direction: $\langle \phi(x,y)\rangle = \phic(y)$, where $x$ is in an element of $M^{(3,1)}$ Minkowski 
space-time, while $y$ is the extra coordinate (hereafter we will restrict ourselves to $R=S^1/Z_2$).
Recently, it has been noticed~\cite{Rubakov:bb, Georgi:2000wb}  
that the non-trivial $y$-dependence of the vev of a scalar field could be very useful phenomenologically,
since it may lead to a natural localization of higher-dimensional fermions on lower-dimensional manifolds
(``fat brane scenario''). 
This is a very attractive approach, since it allows us to control the effective couplings between 4D states
by tuning the overlap of their wave function in the extra dimension~\cite{Kaplan:2001ga, Arkani-Hamed:1999dc}.
In fact a similar idea had already been explored 
in the mid-seventies, in the context of domain walls that separate regions of 
different vacuum expectation values of a scalar field \cite{Voloshin:aa}, see also \cite{Jackiw:1976xx}.

The standard approximation adopted in the studies of non-trivial scalar vev's in extended compact space-times 
was to assume 
that the size of the extra dimension $L$ is
much larger than the typical scale of the scalar field potential $m$. 
In this paper we will show how to construct a 
low-energy effective theory exactly, including the determination of the Kaluza--Klein (KK) 
states of scalars and fermions
and their mass spectrum, in the case of a scalar with a Higgs-like potential in 5D.

First we will consider a 5D model of a free real scalar field $\phi=\phi(x,y)$ defined by the following
Lagrangian density:
\beq
\lcal^{(5)} =\frac12 \; \partial^M \phi(x,y) \;\partial_M \phi(x,y) - \v5 (\phi)
\label{scal_models}
\eeq
with $y\equiv x^4$ and 
\beq
\v5(\phi)= {\lam \over 4}\left(|\phi|^2-{m^2 \over \lam} \right)^2\,.
\label{v5}
\eeq
Below we will look for a field configuration $\phi_c(x,y)$ that corresponds to the extrema of the 
classical action. 
Since the fifth dimension is compact we will allow for a non-trivial dependence of the background 
solution on the extra coordinate.
In addition, we will restrict ourselves to time-independent solutions, so $\phi_c=\phi_c(y)$.

After a single integration of the Euler--Lagrange equation:
\beq
\partial_M \partial^M \phi + {\partial \v5 \over \partial \phi} = 0
\label{scal_el}
\eeq
we obtain the following equation for the background $\phi_c$ 
that describes the energy conservation of the system:
\beq
\frac12 \phi_c^{'\;2}-V(\phi_c)=E={\rm const.} \,,
\label{energy}
\eeq
where $^{'}$ denotes a differentiation with respect to $y$.
It is seen that our field theoretical problem corresponds to a 1D motion of the classical
material point in the potential $-V(\phic)$, where $\phic$ is the coordinate,  $y$ corresponds to 
the time and $E$ is interpreted as the total energy. 
It is easy to see that for $E=0$ we obtain a non-trivial solution of (\ref{energy}) known 
as the kink and antikink:
\beq
\phi_c^{\rm kink}(y;y_0)= \pm {m \over \sqrt{\lam}} \tanh\left[{m \over \sqrt{2}} (y-y_0)\right]
\label{kink}
\eeq
where the constant $y_0$ is the ``kink-location''.
The kink solution corresponds to the motion of the point, starting with zero velocity
at the top of one of the hills of the potential $-V(\phic)$ (located at $\phi_c =\pm m/\sqrt{\lam}$) and moving 
down toward the origin, and then,
after an infinite amount of ``time'' ($y$), reaching the next hill\footnote{
There exists also another solution of Eq.~(\ref{energy})
such that $h(y) \propto \coth\left[{m \over \sqrt{2}} (y-y_0)\right]$. 
This solution corresponds to a particle that starts, at
finite moment $y=y_0$ with an infinite velocity at an infinite distance. Since it is
singular at $y=y_0$ we will not consider it hereafter in the field theoretical 
model we will discuss.}.
The above solution assumes that the integration constant $E$
of Eq.~(\ref{energy}) is zero. It
is easy to argue that this is really a necessary choice 
for a non-compact space\footnote{For a non-zero 
integration constant there would be a constant contribution to $h(y)^{'}$ at infinity, which would therefore
lead to infinite total energy $E=\int_{-\infty}^{+\infty} T_{00}$.}, however in our 
case of the compact fifth dimension there is no reason to assume $E=0$.
From the mechanical analogy it is also seen that
if the material point starts at $y$, it will then move toward the origin and eventually, after 
a finite time, reach the
symmetric position at $-y$. Obviously the point then returns, so that the motion will be periodic\footnote{ 
For the related discussion of instantons within a single scalar field theory in one time and zero space 
dimensions
for the double-well anharmonic oscillator, see Refs. \cite{Harrington:tx} and \cite{Richard:1980ei}. The 
periodic background 
solutions that we discuss here were also considered in Ref. \cite{Carrillo:2000} in the context of 1+1 
$\phi^4$ scalar field theory.}.

The paper is organized as follows. In Sec.~\ref{per_kink} we will 
find a periodic kink-like solution for the scalar background. The 
KK modes and their corresponding masses for the free scalar considered will be
determined there. 
In section~\ref{per_local} we will add a fermion field coupled to the scalar. We will show how to solve 
the Schr\"odinger-like equation for all the fermionic modes and their corresponding masses, 
and in this way we find states not considered until now, which have very interesting implications, 
both phenomenologically and in model building.
We then summarize our results in the final Section~\ref{com}.

\section{Periodic Kink-Like Background Profile}
\label{per_kink}

It is straightforward to integrate Eq. (\ref{energy}) even for $E\neq 0$:
\beq
y-y_0 = \pm {1\over m} \int {d\varphi\over \sqrt{\varphi^4 - \varphi^2 + {c \over 2}  }}\, ,
\eeq
where $y_0$ and $E$ are constants of integration and we have defined the dimensionless parameter 
$c={1\over 2}+ {2 E \lambda \over  m^4}$, and the field has been rescaled as follows: 
$\varphi=\lam/(2 m^2)\phic$. 

We are interested in solutions $\phic(y)$ (or $\varphi(y)$), which are periodic and continuous. It is 
easy to see from the mechanical model that this can only happen when $0<c<1/2$. 
In that case the background  solution is bounded by $|\phic(y)| \leq {m \over \sqrt{\lambda}}$ and is given by
\beq
\hspace{-1cm}\phic(y)=\pm 
{m\over \sqrt{\lambda}} \sqrt{2 k^2 \over k^2+1}\   {\rm sn}\left({m (y-y_0) \over \sqrt{k^2 +1}} ,\ k^2 \right)
\label{periodicvev}
\eeq
where 
\beq
k^2= {1- \sqrt{1-2c} \over c}-1={m^2-2\sqrt{|\lam E|} \over m^2+2\sqrt{|\lam E|} }
\eeq
and ${\rm sn}(x,k^2)$ is the Jacobi elliptic-sine amplitude of modulus $k$ \cite{erdelyi}.
These solutions are parametrized by the modulus $k$, and since  $0<c<1/2$, we have $0<k<1$.
The function ${\rm sn}(x,k^2)$ is odd in $x$, and oscillates between $1$ and $-1$. Its period is $4 K$ where
\bea
K(k^2)=\int_0^{\pi/2} {d\theta \over (1 - k^2 \sin^2\theta)} = F\left({\pi\over2},k\right)
\eea
is the complete elliptic integral of the first kind \cite{erdelyi}.
The modulus $k$ depends on the constant $c$, which in turn depends on $E$, and therefore it is a 
free parameter that will determine the size of the extra dimension $L$ through the value of the period 
$\omega=4\sqrt{k^2+1}K(k^2)/m$~\footnote{Note that
the size of the 
extra dimension $L$ does not need to be equal to the period of the background solution, but 
it must rather be a multiple of the period: $L=n\omega$.}. 
We see that the size of the extra dimension is not fixed by any dynamics, and therefore an unknown 
stabilization mechanism must be assumed.

\begin{figure}[h]
\centering
\includegraphics[width=12cm]{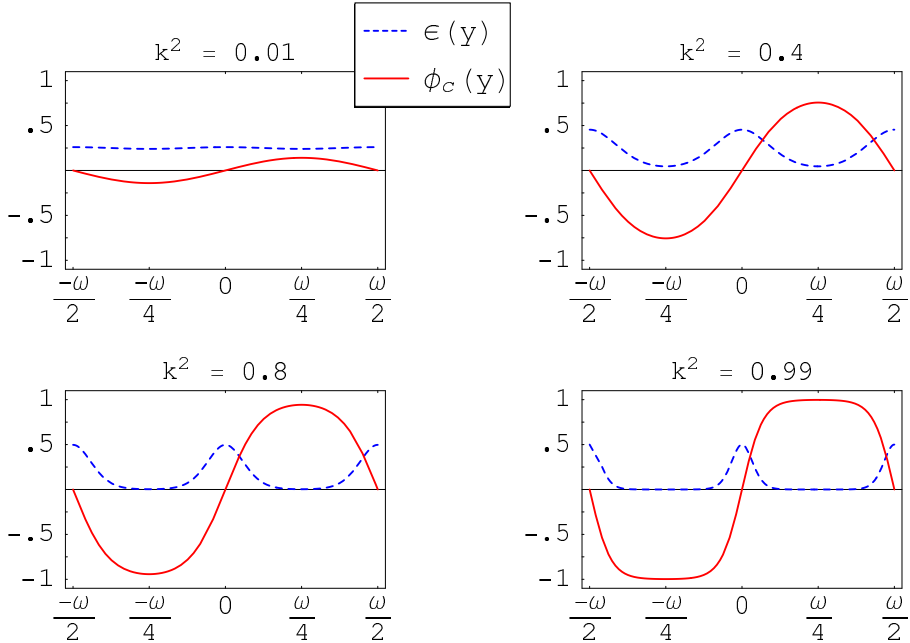}
\caption{The profiles of the non-trivial background $\phic(y)$ (in units of $m/\sqrt{\lambda}$) and 
the ``energy density'' $\eps (y)$ (in units of $m^4/\lambda$) for $k^2=0.01,\ 0.4,\ 0.8$ and $0.99$ 
(or $\omega =1.01\omega_0,\ 1.34\omega_0 ,\ 1.93\omega_0$ and $3.32\omega_0$, with $\omega_0= 2 \pi / m$). 
We have chosen $y_0=0$; for $y_0 \neq 0$, $y$ should be replaced  by $y+y_0$. }
\label{nontrivev}
\end{figure}

In Fig. \ref{nontrivev} we show the profile of the non-trivial background $\phic(y)$ for different values of 
$k^2$ (i.e. for different compactification scales).
We also show the energy density $\eps(y)$
\beq
\eps(y)={1\over 2} (\partial_y \phic )^2 + V(\phic) \,.
\eeq

We can now look at two interesting limits:
\bit
\item $k \to 0$, then $\phic (y) \to  \pm {m\over \sqrt{\lambda}} \sqrt{2} k \ \sin{ m  (y-y_0)} + 
{\cal O}(k^2)\ \ \to 0 $.
The background thus vanishes in the limit.
\item $k \to 1$, then $\phic (y) \to \pm {m\over \sqrt{\lambda}} \tanh{m \sqrt{1 \over 2}\ (y-y_0)} $. 
Note that in this
limit the periodicity of the background is being lost,  even though $\phic (y)$ is a perfectly 
periodic function for any $k<1$.
\eit
The first limit corresponds to $c \to 0$. In this case although the non-trivial profile vanishes, 
there is a limiting 
period $\omega_0\equiv\omega\left.\right|_{k=0}= 4 K(0)/m =2\pi/ m$. This means that there is a minimum 
size for the extra dimension, which is fixed by the 
mass parameter $m$ of the potential. 
This is easy to understand by recalling our mechanical analogue 
of the field theoretical problem. Namely, the limit
$k\to 0$ ($c\to 0$) corresponds to the total energy $E\to -m^4/(4 \lam)$ that is the bottom of the
classical potential ($-V(h)$). In that region, since the quartic part of the potential could be neglected, 
we observe small 
harmonic oscillations around the minimum $y=0$~\footnote{Hereafter we will choose $y_0=0$.}. 
There thus exists a minimal period that is fixed to be $2 \pi /m$
by the quadratic term (the mass term) of the potential. Consequently, {\it if the
scalar background is non-trivial, then there is
a minimal size of the extra dimension that is a multiple of $\omega_0$}. Note that we did not to impose
the periodicity; whenever $0<k<1$ the solution must be periodic for any parameter of the model, no 
fine-tuning is needed.

The second limit, $k\to 1$ ($c\to 1/2$) corresponds to $E\to 0$, i.e. the motion that starts exactly at the
top of a hill of the classical potential ($-V(h)$). In this case the total available time ($y$) 
needed to reach the other summit is infinite and therefore so is the corresponding size of the extra dimension 
$L$.

\subsection{Scalar spectrum}
\label{Scalar_spectrum}

\begin{table}[b]
\hspace{-1.2cm}\begin{tabular}{||c|c|c|c|c|c|c||} 
\hline
$n$  & $ \phi_n   $ & $\frac{M_n^2}{m^2}$ &  $\left.\phi_n\right|_{k^2=1}$ & 
$\left.\frac{M_n^2}{m^2}\right|_{k^2=1}$ & $\left.\frac{M_n^2}{m^2}\right|_{k^2=0}$& Period  \\ \hline
$0^{++} $ & ${\rm sn}^2(s,k^2) - {1+k^2+\sqrt{1- k^2+k^4}\over 3 k^2}$ & $ 1-\sqrt{1+3 ({1-k^2\over 1+k^2})^2}$ 
& ${\rm tanh}^2(s)-1$ & $0$ & $-1$& $\omega' /2$         \\ 
$0^{+-} $&${\rm cn}(s,k^2)\ {\rm dn}(s,k^2)  $  & 0 & ${\rm sech}^2(s)$ & $0$ & $0$& $\omega'$ \\ 
$1^{-+}\ $ & ${\rm sn}(s,k^2)\ {\rm dn}(s,k^2)  $ &  $  {3k^2\over 1+k^2}$  & ${\rm sech}(s)\ {\rm tanh}(s)$ & 
$\frac{3}{2}$ & $0$& $\omega'$ \\ 
$1^{--}$ & ${\rm sn}(s,k^2)\ {\rm cn}(s,k^2)  $  &  ${3\over 1+k^2} $ & ${\rm sech}(s)\ {\rm tanh}(s)$ & 
$\frac{3}{2}$ & $3$ & $\omega' /2$ \\ 
$2^{++}\ $& ${\rm sn}^2(s,k^2) - {1+k^2-\sqrt{1- k^2+k^4}\over 3 k^2}$&$1+\sqrt{1+3 ({1-k^2\over 1+k^2})^2}$& 
${\rm tanh}^2(s)-\frac13$ &
 2 & $3$ & $\omega' /2$ \\ \hline
\end{tabular}
\caption{The first five unnormalized scalar eigenfunctions and the resulting mass spectrum from Eq. 
(\ref{philameeq}). Here, $\omega'=4K(k^2)$ is the ``rescaled'' period of the background solution $\phic(s)$.
The parities are defined with respect to $s=0$ and $s=\omega'/4$.}
\label{lamepol}
\end{table}

Let us now expand the Lagrangian (\ref{scal_models}) around the background (\ref{periodicvev}):
\beq 
\phi(x,y)\to \phic(y) + \phi(x,y) \,,
\eeq
where $\phi(x,y)$ that appears on the right-hand side of $\to$ is a fluctuation.
Let us separate contributions to the Lagrangian density that are quadratic in the fluctuations $\phi(x,y)$:
\beq
\lcal(x,y)=\frac12 \partial_\mu \phi \partial^\mu \phi - \frac12 \; \phi\left[-{d^2 \over dy^2} +
\left.{\partial^2\v5 \over \partial \phi^2}\right|_{\phi=\phic}\right]\phi + \cdots\,,
\eeq
where I have integrated by parts and dropped the total derivative because of periodicity. 
Let us expand in terms of the KK modes that are defined in such a way that the mass matrix is diagonal:
\beq
\left[-{d^2 \over dy^2} + \left.{\partial^2\v5 \over \partial \phi^2}\right|_{\phi=\phic}\right] \phi_n(y)=
M_n^2 \phi_n(y),
\label{seq}
\eeq
where the expansion $\phi(x,y)=\sum \phi_n(y)\chi_n(x)$ was adopted, and $M_n$ denote diagonal masses of the KK
modes $\chi_n(x)$.

Substituting the background $\phic(y)$ from Eq.~(\ref{periodicvev}) we 
obtain the following equation to determine the right KK basis:
\bea
 -\phi_n''(y) + \left[6  m^2  {k^2 \over k^2+1}   {\rm sn}^2\left(m \sqrt{1 \over k^2 +1}\ (y-y_0),\ k^2\right) 
- 
m^2 -M_n^2\right]\  \phi_n(y) = 0 
\eea
By a suitable rescaling, $s = {m \over \sqrt{k^2+1}}\ (y-y_0)$ we can get
\bea
 \phi_n''(s) + \left[ (k^2 +1) (1 + {M_n^2\over m^2}) -6 k^2  {\rm sn}^2\!\left(s,\ k^2\right) \right]\  
\phi_n(s) = 0 \,,
\label{philameeq}
\eea
which is the Lam\'e equation (in the Jacobian form)
\beq
 \phi_n''(s) + \left[ h -n (n+1) k^2  {\rm sn}^2\!\left(s,\ k^2\right) \right]\  \phi_n(s) = 0 \,,
\label{lame}
\eeq
where $n=2$ and $h=  (k^2 +1) (1 + {M_n^2\over m^2})$.
It is known that for given $n$ and $k$, periodic solutions of the Lam\'e equation exist for an infinite sequence of 
characteristic values of $h$ \cite{erdelyi}. Moreover, the periodic solutions will be of period $2K(k^2)$ 
and $4K(k^2)$, where $2K(k^2)$ is the period of the Lam\'e equation (\ref{lame}). 

Each of these characteristic values of $h$ will determine the KK mass of the corresponding mode.
In the case that $n$ in (\ref{lame}) is integer, the first $2n+1$ solutions will be polynomials in the 
Jacobi elliptic functions. 
For the case $n=2$ at hand, the first 5 eigenmodes are simple enough \cite{Harrington:tx,Arscott} and are 
given in 
Table~\ref{lamepol}, along with their eigenvalues and parities along the extra-dimension.
The rest of the spectrum is given by transcendental Lam\'e functions, which can be expanded in infinite series 
(of trigonometric functions or Legendre functions).

As is seen from Table~\ref{lamepol} the lowest eigenvalue of the quadratic operator 
$-d^2/dy^2 + \partial^2\v5/ \partial \phi^2_{|\phi=\phic}$ is negative\footnote{Since the background
solution $\phic$ satisfies  Eq.~(\ref{energy}), its derivative must be a zero-mode
wave function; see Table~\ref{lamepol} for the eigenfunction 
$\phi_{0^{+-}}\propto d\;{\rm sn}(s,k^2)/ds={\rm cn}(s,k^2)\ {\rm dn}(s,k^2)$. 
The derivative must have at least one node (as a consequence of the
periodicity of $\phic$) so that, on the virtue of Sturm's theorem, there must exist a state of 
lower energy, i.e. negative $M_n^2$. This is what is being observed in Table~\ref{lamepol} 
for the state $n=0^{++}$.}; 
our background solution (\ref{periodicvev}) is therefore unstable in time evolution\cite{col}. Note, 
however, that the ground state (which causes the instability problem) 
is even under $s\to -s$ (as it should since it must have no nodes).
Therefore the problem could be solved by extending the antisymmetry of the background 
solution\footnote{Note that the requirement of the antisymmetry explains the issue of the natural
existence of the non-trivial background. Without the antisymmetry, the trivial
$\phic (y)=0$ solution that corresponds to lower total energy would be preferred energetically and 
would therefore determine the real ground state of the system.}
$\phi_c$, also for the KK excitations. So, we will require the following orbifold boundary 
condition\footnote{For $y_0\neq 0$ one should replace 
(\ref{boundas}) with 
$\phi(y_0+y)=-\phi(y_0-y)$.}:
\beq
\phi(y)=-\phi(-y)
\label{boundas}
\eeq
as they eliminate all the even modes leaving only $n=1^{-+}$ and $n=1^{--}$ in the first 
5 modes.
Note that $\phi(L/2+y)=-\phi(L/2-y)$ follows from (\ref{boundas}), as a consequence of periodicity.
\begin{figure}[h]
\hspace{-.1cm}\includegraphics[width=15cm]{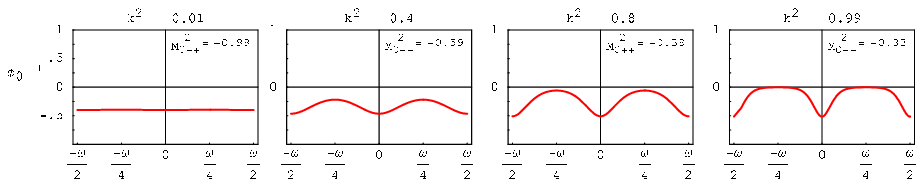}
\includegraphics[width=15cm]{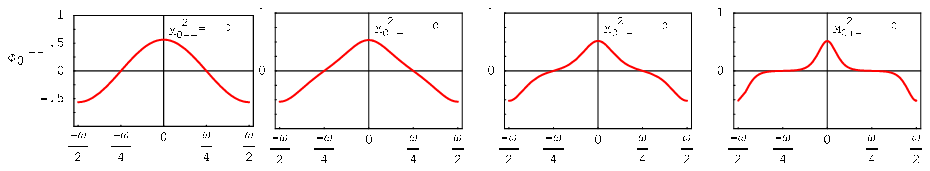}  
\includegraphics[width=15cm]{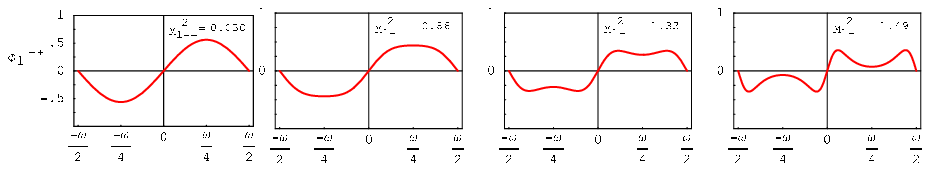}
\includegraphics[width=15cm]{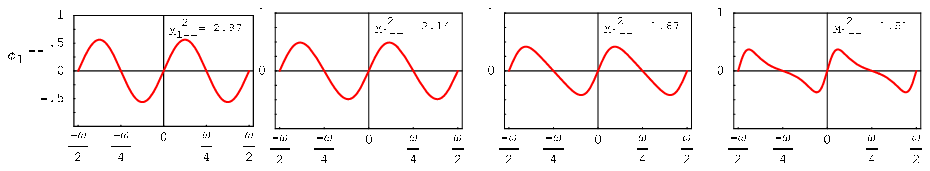}
\includegraphics[width=15cm]{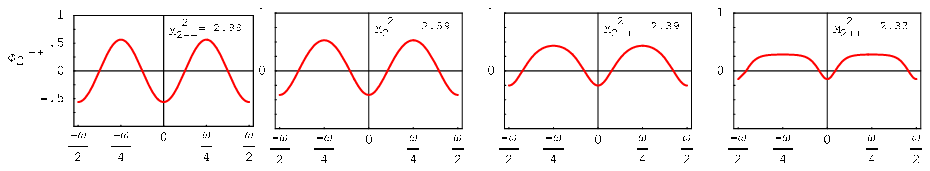}
\caption{The normalized profiles of the first scalar modes  
for $k^2=0.01,\ 0.4,\ 0.8$ and $0.99$ (or $\omega =1.01\omega_0,\ 1.34\omega_0 ,\ 1.93\omega_0$ and 
$3.32\omega_0$, with $\omega_0= 2 \pi / m$). The mass eigenvalues are given in units of $m^2$. 
We have chosen $y_0=0$; for $y_0 \neq 0$, $y$ one should be replaced by $y+y_0$.}
\label{philames}
\end{figure}
It is worth emphasizing that the antisymmetry is also essential to develop a non-trivial vev: if
symmetric solutions were allowed, then the {\it constant} vev would be chosen, as it is energetically 
more convenient.

\section{Fermion Localization}
\label{per_local}


A non-trivial vacuum of the scalar field will influence the phenomenology of all fields that couple 
to the scalar.
Here we will consider the simplest scenario of the real scalar field coupled to fermions defined by the 
following
Lagrangian density:
\bea
\lcal&=&\bar{\psi}(x,y)\left[i\Gamma^M\partial_M - f \phi(x,y) - m_0\right]\psi(x,y) \\
&&+ \frac12 \; \partial^M \phi(x,y) \;\partial_M \phi(x,y)
- {\lam \over 4}\left(|\phi|^2-{m^2 \over \lam} \right)^2\non \,,
\label{model1bis}
\eea
with $\Gamma^5=i\gamma_5$. For simplicity, in the following discussion 
we will consider a massless 5D fermion,
so that at the end we obtain chiral fermionic fields whose masses are 
generated exclusively through spontaneous symmetry breaking in the presence of
Yukawa couplings. The absence of the mass term will be guaranteed by requiring the action to be 
symmetric under the following transformations:
\bea
\phi(x,y)\to -\phi(x, L/2 - y)\lsp {\rm and} \lsp \psi(x,y)\to \gamma_5 \psi(x, L/2 - y)
\label{symmetry}
\eea
We will adopt the following orbifold boundary conditions:
\beq
\psi(y)=\gamma_5\psi(-y) \lsp {\rm and} \lsp \psi(L/2+y)=\gamma_5\psi(L/2-y)\, ,
\label{fer_boud}
\eeq 
so that right- and left-handed fermions are symmetric and antisymmetric with respect to $y=0$ and 
$y=L/2$ (which is identified with $y=-L/2$)~\footnote{Note that the second condition of Eq.~(\ref{fer_boud}) 
is a consequence of the first one and of the requirement of periodicity.}. It is crucial to construct a chiral
theory that disentangles left- and right-handed 4D fermions.

As we will see, once the field $\phi(x,y)$ acquires its non-trivial vev $\phic$ (see Eq.~(\ref{periodicvev})), 
the mass spectrum of the KK modes of the fermion field is altered, from the ``usual'' tower of 
fermions with masses $m_n=2 \pi  n/L$. To find the spectrum and the eigenfunctions (KK modes)
let us first decompose the field $\psi(x,y)$ in chiral components:
\bea
\psi&=& \psi_{\rm L} + \psi_{\rm R}
\eea
and let us now write the coupled Weyl equations of motion for these components:
\bea
&&i \partial\hspace{-.21cm}/\   \psi_{\rm L}(x,y)  - (\partial_5 +  f \phic(y)) \psi_{\rm R}(x,y) = 0\\
&&i \partial \hspace{-.21cm}/\   \psi_{\rm R}(x,y) + (\partial_5 -  f \phic(y)) \psi_{\rm L}(x,y) = 0\,.
\eea
Now, we want to separate variables for each chiral field
\bea
\psi_{\rm L}(x,y)&=&\psi_x^{\rm L}(x)\ \psi^{\rm L}_y(y)\\
\psi_{\rm R}(x,y)&=&\psi_x^{\rm R}(x)\ \psi^{\rm R}_y(y)\,,
\eea
which gives
\bea
&i \partial \hspace{-.21cm}/\   \psi_x^{\rm L} - m_{n} \psi_x^{\rm R}=0,&\ \     (\partial_y  +   f \phic(y)) \psi^{\rm R}_y =  
m_n \psi^{\rm L}_y\label{RL1} \\
&i \partial \hspace{-.21cm}/\   \psi_x^{\rm R} - m_{n} \psi_x^{\rm L}=0,& \ \    (-\partial_y  +   f \phic(y))  \psi^{\rm L}_y = 
m_n \psi^{\rm R}_y\,.
\label{LR1}
\eea
To obtain separate equations for the functions $\psi^{\rm L}_y$ and $\psi^{\rm R}_y$ we apply the operator  
$(-\partial_5+f \phic)$ 
to the second equation in (\ref{RL1}) and $( \partial_5 + f \phic)$ to the second one in (\ref{LR1}), 
and get 
\bea
\left[-\partial_y^2 + \left(f \phic\right)^2 \mp f \phic'- m^2_{n}\right] \psi^{\rm R/L}_y = 0\non\\
{\rm or}\hspace{3cm} \ &&\non\\
\left[-\partial_y^2 + V_{\rm R/L}- m^2_{n} \right] \psi^{\rm R/L}_y = 0
\label{RLy}
\eea
In general, since left- and right-handed profiles are solutions of different equations, we can expect 
different properties of 
$\psi^{\rm L}_y$ and $\psi_y^{\rm R}$, so that the theory is chiral, as we demanded.
The ``difference'' between chiralities is controlled by the Yukawa coupling $f$ inside the 
potentials 
\beq
V_{\rm R/L} \equiv \left(f \phic\right)^2 \mp f \phic^{'} =
(\delta\ k)^2\ {\rm sn}^2\!\left(s,k^2\right) \mp (\delta\ k)\ {\rm cn}\!\left(s,k^2\right)\
 {\rm dn}\!\left(s,k^2\right)\,.
\label{vlr}
\eeq
Equations (\ref{RLy}) have the form of the Hill's equation, and it is known that if $P$ is 
the period of the equation, there will exist periodic solutions, of period $P$ or $2P$, and of defined even
or odd parity. For each of these solutions there will be a specific eigenvalue $(m^{\rm R/L}_n)^2$. 
In our case the required degeneracy of left- and right-handed modes will be guaranteed for all 
solutions of both equations. 
If the periodic background vev has some extra symmetry, it is possible to address this issue before 
actually solving Eqs.~(\ref{RLy}).

\begin{figure}[h]

\hspace{-.8cm}\includegraphics[width=17cm]{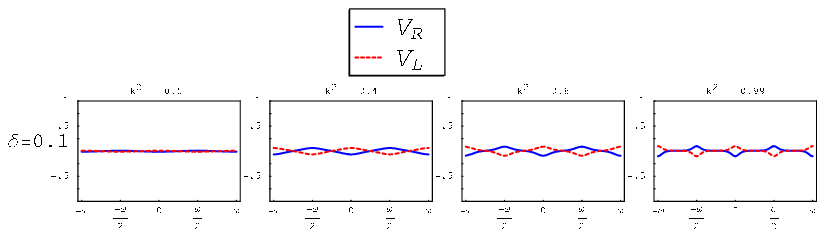}

\hspace{-1.2cm} \includegraphics[width=17.2cm]{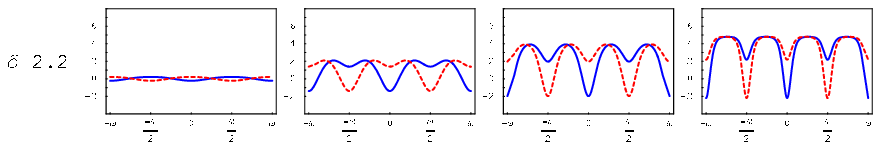}

\hspace{-1.45cm} \includegraphics[width=17.4cm]{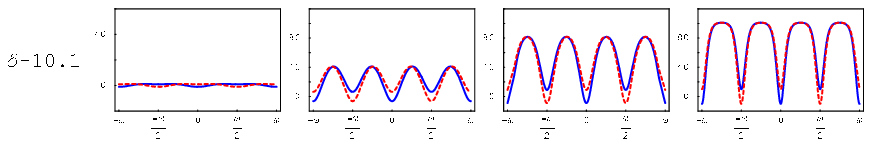}
\caption{The profiles of the potentials $V_{\rm R}$ and $V_{\rm L}$ for $k^2=0.01,\ 0.4,\ 0.8$ and $0.99$ 
(or $\omega =1.01\omega_0,\ 1.34\omega_0 ,\ 1.93\omega_0$ and $3.32\omega_0$, with $\omega_0= 2 \pi / m$) 
and for $\delta=0.1,\ 1$ 
and $10$. We have chosen $y_0=0$; for $y_0 \neq 0$, $y$ should be replaced by $y+y_0$. (Note the scale changes.)}
 \label{VRVL}
 \end{figure}

Let the background vev $\phic$ be periodic, of period $\omega$, i.e. $\ \phic(y + \omega)=\phic(y)$.
Assume that this background vev also satisfies the relation \bea\phic(y + \omega/2)= -\phic(y) \label{P2sym}\eea
If this is the case, we immediately see that 
\bea
V_{\rm R}(y+\omega/2) &=& V_{\rm L}(y) \,.
\label{transsym}
\eea
This means that all periodic solutions of both equations will have exactly the same spectrum, the solutions 
being just shifted by $\omega/2$.
Therefore it is sufficient to solve only one of the equations (R or L) and find the corresponding
spectrum. The translational symmetry (\ref{transsym}) will then immediately give the general solutions 
(with the same spectrum) to the other equation.
With the general solutions of both equations, one can then impose the boundary conditions we have chosen 
for each chiral fermion, and get the final solutions for the fermion modes.
It turns out that the solution of $\phic$ from Eq. (\ref{periodicvev}) satisfies the relation (\ref{P2sym}). 
In Fig.~\ref{VRVL} we plot the potentials $V_{\rm R}$ and $V_{\rm L}$ for selected values of $k^2$ (that
determines the period $\omega$). 

We will now solve the R equation (and therefore get also the L solutions, thanks to the translational 
symmetry) and show the main results here. We will refer the reader to the Appendix for a 
more detailed derivation.

\begin{figure}[h]

\hspace{-.2cm}\includegraphics[width=16.5cm]{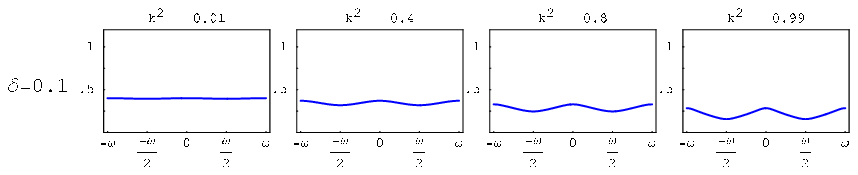}

\hspace{-.3cm}\includegraphics[width=16.4cm]{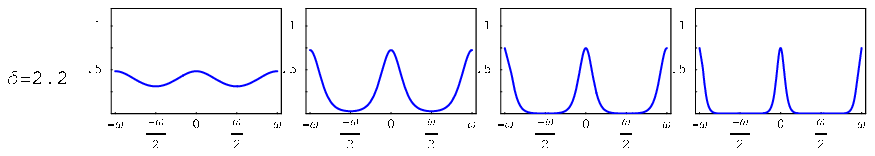}

\hspace{-.65cm}\includegraphics[width=16.8cm]{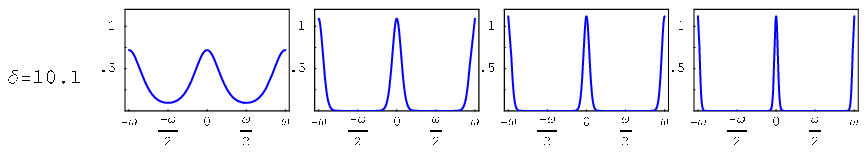}
\caption{The profile of the normalized zero mode $\psi_0^{\rm R}$ for $k^2=0.01,\ 0.4,\ 0.8$ 
and $0.99$ (or $\omega =1.01\omega_0,\ 1.34\omega_0 ,\ 1.93\omega_0$ and $3.32\omega_0$, with 
$\omega_0= 2 \pi / m$) 
and with $\delta=0.1,\ 2.2$ and $10.1$. We have chosen $y_0=0$; for $y_0 \neq 0$, $y$ should be replaced 
by $y+y_0$.}
\label{fzeromodels}
\end{figure}

Using the background solution $\phic$ from Eq.~(\ref{periodicvev}) one can find that the unnormalized chiral 
zero mode consistent with the boundary condition (\ref{fer_boud}) has quite a simple form:
\bea
\psi_0^{\rm R}(y) \propto  \left[ {\rm dn}\!\left({m \over \sqrt{k^2+1}}\ y ,\ k^2\right) 
+ k\ {\rm cn}\!\left({m \over \sqrt{k^2+1}}\ y ,\ k^2\right) \right]^\delta \;\; {\rm and} \;\; 
\psi_0^{\rm L}(y) =0\,,
\label{psiRzero}
\eea
where $\delta = \sqrt{2 f^2 / \lambda}$. 
This solution is exact and we can see the behaviour of this chiral mode for different compactification scales 
(i.e. $k^2$) and for different values of $\delta$ in Fig.~\ref{fzeromodels}. The figure illustrates the
relevance of the Yukawa coupling ($\propto \delta$) for the efficiency of the localization of the 
fermionic KK modes. 
It is also seen that it is easier for a given strength of the Yukawa coupling,  to 
obtain the localization for the large extra dimension (and thus for large $k^2$). 
The rest of the modes cannot be written in such a compact form as the zero mode; however,
it can be given in terms of convergent infinite trigonometric series. 
Their spectrum can be found by numerically solving a transcendental equation for each mass eigenvalue.
Once the boundary conditions are imposed, we have
\bea
\psi_{2n}^{\rm R}(y)   &\propto& \psi_0(y)\ \ CI^{2n}_\omega\!\!\left[\ {\rm am}\!\left({m\over 2} {(1+k)\over 
\sqrt{k^2+1}}\ y,\ {4 k \over (1+k)^2} \right)\right]\\
\psi_{2n+1}^{\rm R}(y) &\propto& \psi_0(y)\ \ CI^{2n+1}_{2\omega}\!\!\left[\ {\rm am}\!\left({m\over 2} {(1+k)\over 
\sqrt{k^2+1}}\ y,\ {4 k \over (1+k)^2} \right)\right]  \\
\psi_{2n}^{\rm L}(y)   &\propto& {\psi_0(y+\omega/2)}\ \ SI^{2n}_\omega\!\!\left[\ {\rm am}\!\left({m\over 2} {(1+k)\over 
\sqrt{k^2+1}} (y+\omega/2),\ {4 k \over (1+k)^2} \right)\right] \\
\psi_{2n+1}^{\rm L}(y) &\propto& {\psi_0(y+\omega/2)}\ \  CI^{2n+1}_{2\omega}\!\!\left[\ {\rm am}\!\left({m\over 2} 
{(1+k)\over  \sqrt{k^2+1}} (y+\omega/2),\ {4 k \over (1+k)^2} \right)\right]\,,
\eea
where
\beq
\psi_0(y)=\left[ {\rm dn}\!\left({m \over \sqrt{k^2+1}}\ y ,\ k^2\right) 
+ k\ {\rm cn}\!\left({m \over \sqrt{k^2+1}}\ y ,\ k^2\right) \right]^\delta\
\eeq
and $\omega= {4\sqrt{1+k^2}\over m} K(k^2) $ is the period of $\phic(y)$. Finally, ${\rm am}\!\left(x,m\right)$ 
is the Jacobian elliptic amplitude. 

The series $CI^{2n}_\pi(t)$, $SI^{2n}_\pi(t)$ and $CI^{2n+1}_{2\pi}(t)$ are the periodic solutions 
(of period $\pi$ and $2\pi$ in the $t$ variable) of the generalized Ince equation~\cite{Magnus} 
(it is a four-parameter, second-order differential equation, but in our case the parameter $d$ in 
\cite{Magnus} is zero, making the equation slightly simpler; see the Appendix for more details). 
The $C$ and $S$ 
indicate the parity of the series ($C$-even and $S$-odd). These generalized Ince solutions are given by:
\bea
CI^{2n}_\pi(t)&=&\sum^{\infty}_{r=0} A^{2n}_{2r} \cos{2r t}\\ 
SI^{2n}_\pi(t)&=&\sum^{\infty}_{r=1} B^{2n}_{2r} \sin{2r t}\\ 
CI^{2n+1}_{2\pi}(t)&=&\sum^{\infty}_{r=0} C^{2n+1}_{2r+1} \cos{(2r+1)t} \,,
\eea
where $n=1,2,3,\dots$.
The coefficients  $A^{2n}_{2r}$, $B^{2n}_{2r}$ and $C^{2n+1}_{2r+1}$ satisfy  three-term recurrence 
relations, as given in the Appendix. It could be shown that $\psi_{2n}^{\rm R/L}(y)$ and $\psi_{2n+1}^{\rm R/L}(y)$
have periods $\omega$ and $2\omega$, respectively.
We remind the reader that the 
scalar modes also had two periods, of $\omega/2$ and $\omega$, meaning that together, 
the wave functions of scalars 
and fermions have a total of three different periodicities.

As for the fermionic mass eigenvalues, there is a potential problem for the solutions of period $\omega$. 
The even right handed solutions  ($CI^{2n}_\omega$), when shifted by $\omega/2$, give even solutions, 
and therefore cannot be the odd left-handed modes.
To have the odd left-handed modes, we need to adopt the odd right-handed solutions ($SI^{2n}_\omega$) 
and shift them by $\omega/2$. But a priori these are two different solutions, and they
are not necessarily degenerate, 
as they should since we are solving for the left and right chiral components of the same mode.
However, it can be proved \cite{Magnus} that the two different solutions 
$CI^{2n}_\omega$ and $SI^{2n}_\omega$ of 
our particular generalized Ince equation have same eigenvalues (as needed) because the transcendental 
equations for $A ^{2n}_{2r}$ and $B^{2n}_{2r}$, which set 
the eigenvalues, are the same (see the Appendix), even though the two solutions are different. 
Therefore the two chiral modes $\psi_{2n}^{\rm R}(y)$ and $\psi_{2n}^{\rm L}(y)$ will have the 
same mass eigenvalue.

In the case of the solutions of period $2\omega$, the degeneracy of the two chiral components is guaranteed 
since they are exactly the same solution, but shifted by $\omega/2$ (they have opposite parities 
because the period is $2\omega$, and therefore the parity changes when shifting them by $\omega/2$). A different 
transcendental equation for $C^{2n+1}_{2r+1}$ will set the masses of these $2\omega$ 
modes $\psi_{2n+1}^{\rm R/L}(y)$. 

\begin{figure}[h]
\centering
\includegraphics[width=16.5cm]{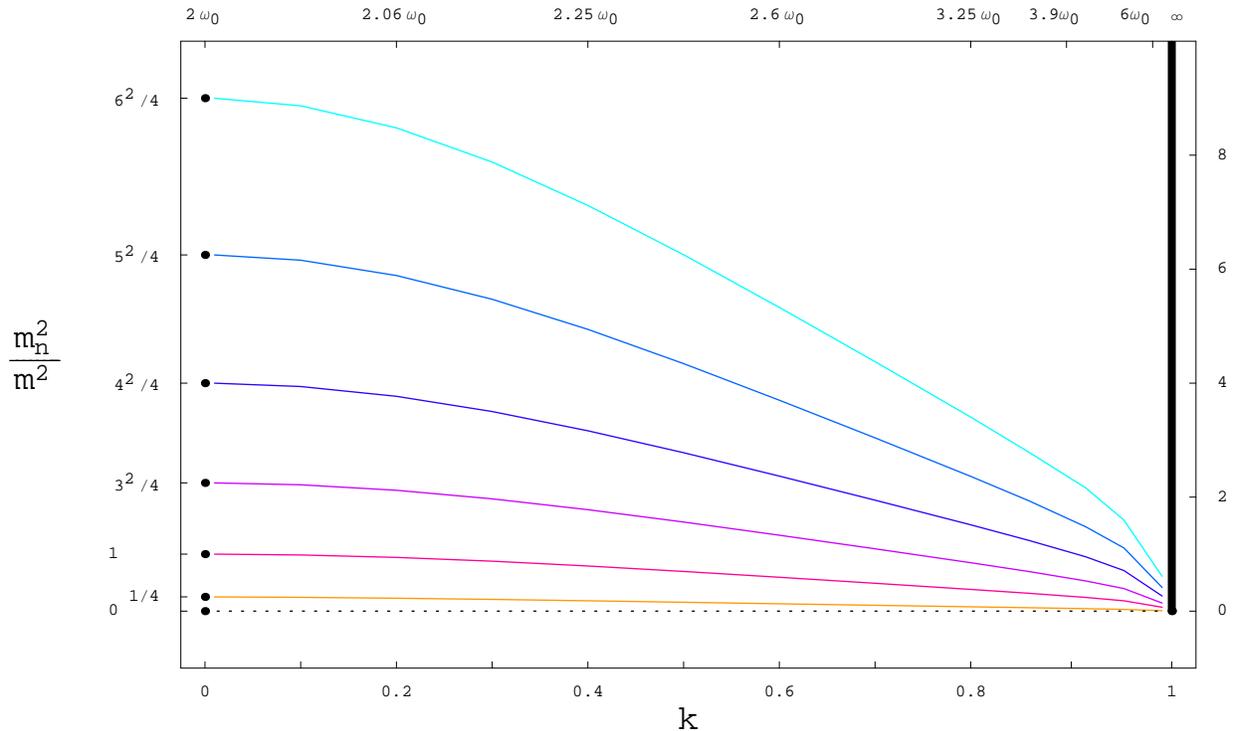}
\caption{The mass variation of the first 7 modes with respect to the size of the extra dimension (i.e. the 
value of $k$). 
The spectrum goes from the discrete $n^2$ type spectrum at $k=0$ to the continuum spectrum for the 
case $k=1$, represented by the thick black vertical line. Here we have taken $\delta=0.1$}
\label{masvsk01}
\end{figure}

\begin{figure}[h]
\centering
\includegraphics[width=17.5cm]{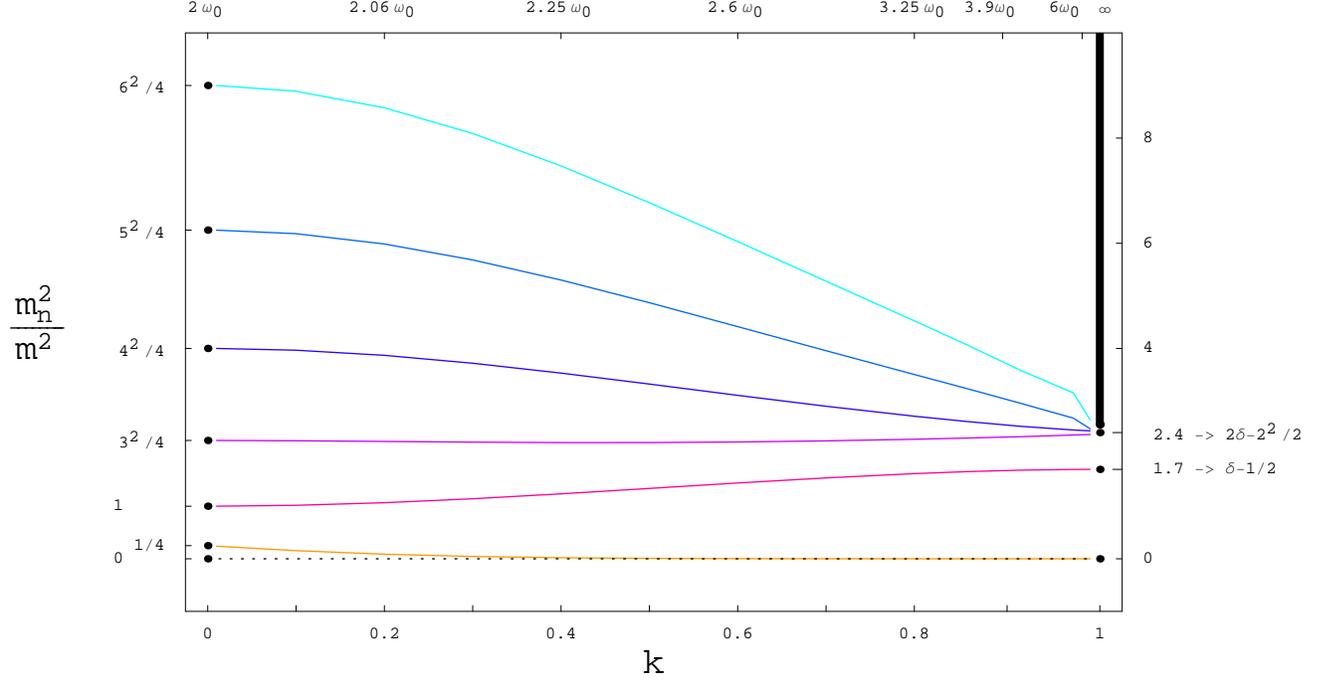}
\vspace{-1.2cm}
\caption{Same as the previous figure, but for $\delta=2.2$. Some discrete modes survive in the $k=1$ limit.}
\label{masvsk22}
\end{figure}

\begin{figure}[h]
\centering
\includegraphics[width=17cm]{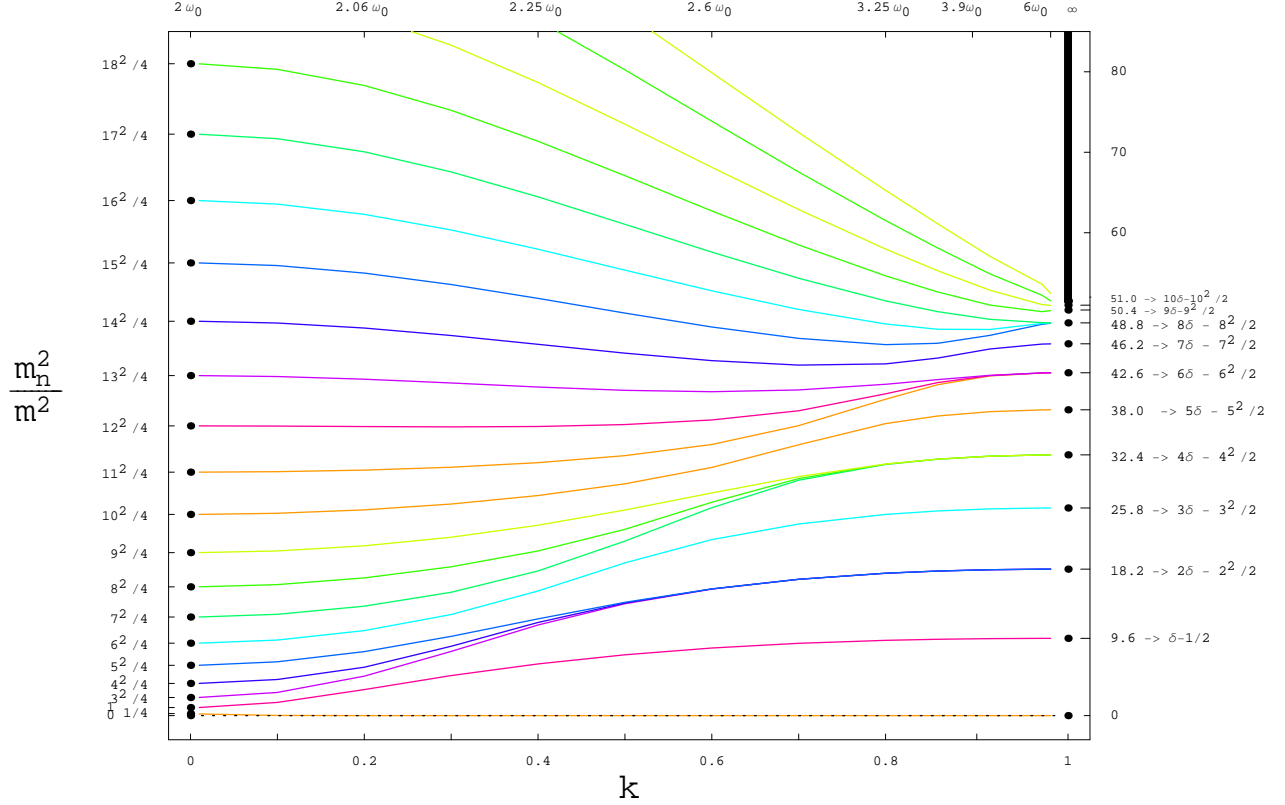}
\vspace{-1.2cm}
\caption{Same as the previous figure, but for the first 22 modes and with $\delta=10.1$. Some more discrete modes survive in the $k=1$ limit.}\label{masvsk51}
\end{figure}

The complete spectrum can be very easily solved numerically, and 
Figs.~\ref{masvsk01}, \ref{masvsk22} and \ref{masvsk51}
show the spectrum of the first fermion modes as a function of $k$, and for three different values of $\delta$.

Before going further, it is instructive to consider the two limiting cases ($k\to 0$) and ($k\to 1$), 
which have known solutions and spectra.
First let us define $s= {m\over \sqrt{k^2+1}}\ y$, so that 
Eqs.~(\ref{RLy}) read
\beq
  \left[-\frac{d^2}{ds^2} + (\delta\ k)^2\ {\rm sn}^2\!\left(s,k^2\right)   \mp 
  (\delta\ k)\ {\rm cn}\!\left(s,k^2\right)\ {\rm dn}\!\left(s,k^2\right)- 
  (k^2+1){m^2_{n}\over m^2} \right]\psi^{\rm R/L}_s = 0\,,  
  \label{Rs}
\eeq
where ${\rm cn}(s,k^2)$ and ${\rm dn}(s,k^2)$ are the Jacobi elliptic 
cosine-amplitude and delta-amplitude respectively. 
 \begin{enumerate}
 \item Keeping only the leading terms in the limit $k\to0$, the equation simplifies to
 \beq
 \left(\frac{d^2}{ds^2} + {m^2_n\over m^2} \right) \psi^{\rm R/L}_s = 0\,,  
\label{Rk0}
 \eeq
with $s=m y$.
In this limit we obtain the following solutions (consistent with the boundary conditions (\ref{fer_boud}))
for the fermionic KK modes:
 \beq
 \psi_n^{\rm R}(y)\propto \cos \left(n\; {\pi\over \omega_0}y\right) \lsp {\rm and} \lsp 
 \psi_n^{\rm L}(y)\propto \sin \left(n\; {\pi\over \omega_0}y\right)\,,
 \label{oscyl}
 \eeq 
with $\omega_0=2\pi/m$~\footnote{Remember that when ($k\to 0$), the size of the extra dimension 
does not vanish, but is instead completely fixed by the quadratic part of the scalar potential.}.
Note that, in the limit $k\to 0$, the Schr\"odinger-like equation is not a periodic equation, and we therefore
have to fix the period by hand; 
here we have assumed that the size of the extra dimension is $2\omega_0$, as this matches the
general solution of the general equation (\ref{RLy}).
The spectrum we expect is therefore $m_n^2=n^2\; {m^2\over 4\ }$ for $n=0,1,2,\dots$, and 
Figs.~\ref{masvsk01}, \ref{masvsk22} and \ref{masvsk51} show how our solutions do approach 
these limiting values.
Note that in this case, even though the theory is chiral, there is no localization of fermions 
observed.
\item When ($k\to1$) (which means $\omega\to\infty$), we have $\ {\rm sn}^2(s,k^2) \to \tanh^2{s}\ $ 
and $\ {\rm cn}(s,k^2) \ {\rm dn}(s,k^2) \to {\rm sech}^2(x)$.

With this limiting behaviour, Eqs.~(\ref{Rs}) become
  \bea
  \left(-\frac{d^2}{ds^2} + \delta^2 \tanh^2{s} \mp \delta\ {\rm sech}^2{s}\ - 2{m^2_n\over m^2} 
\right) \psi^{\rm R/L}_s &=& 0\,,  
  \label{Rk1}
  \eea
for $s=m/\sqrt{2}\ y$.
This is the limit discussed so far in the literature, see \cite{Georgi:2000wb} and 
\cite{Hung:2003cj}.
Since the $y$-direction is no longer compact, we obtain 
in this case both the discrete and the continuum spectrum of eigenvalues. This could be
observed in Figs.~\ref{masvsk01}, \ref{masvsk22} and \ref{masvsk51} where the thick vertical line
represents the continuum spectrum while the discrete eigenvalues are showed as separated 
``energy'' levels. 

Using SUSY quantum mechanics methods, it was shown in \cite{Georgi:2000wb} that when the 
eigenvalue $m_n$ is such that $m_n^2<\delta^2$ (in those authors notation $M^2<w^2$) the first $n_B$ 
fermion modes (with $0\leq n_B <\delta$) are bound states with a very simple discrete spectrum 
given by $ m^2_n = (n_B\ \delta - n_B^2/2)$.\footnote{See also \cite{Hung:2003cj}.}
When $m_n^2>\delta^2$ the spectrum becomes a continuum of states.
Again, Figs.~\ref{masvsk01}, \ref{masvsk22} and \ref{masvsk51} show how this limiting behaviour is 
recovered by our solutions.

\end{enumerate}

\begin{figure}[h]
\hspace{-1.8cm}
\includegraphics[width=19.7cm]{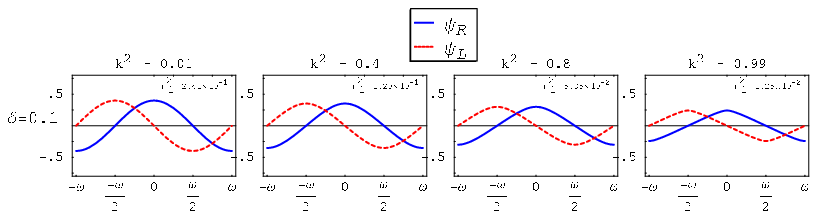}

\hspace{-2cm}
\includegraphics[width=20cm]{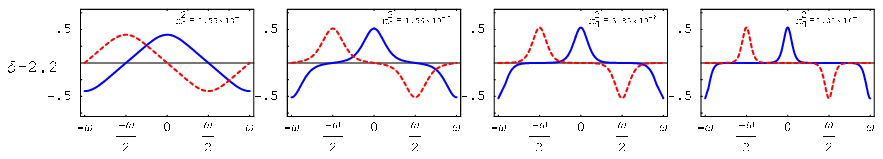}

\hspace{-2cm}
\includegraphics[width=20cm]{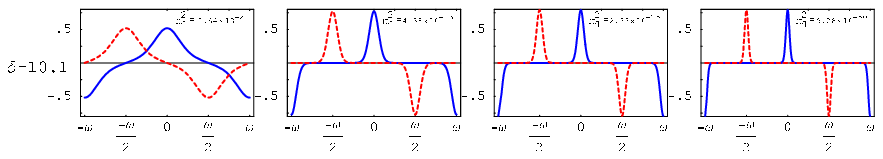}
\caption{The profile of the normalized mode $\psi^{\rm R/L}_1$, of period $2\omega$, for $k^2=0.01,\ 0.4,\ 0.8$ 
and $0.99$ (or $\omega =1.01\omega_0,\ 1.34\omega_0 ,\ 1.93\omega_0$ and $3.32\omega_0$, with $\omega_0= 2 \pi / m$) 
and with $\delta=0.1,\ 2.2$ and $10.1$. The mass eigenvalues are given in units of $m^2$. }
\label{psi1models}
\end{figure}

The early attempts to pass from the case ($k=1$) (infinite extra dimension) to the compact case
($k<1$) seem to have missed half the available spectrum of fermionic modes. The reason for this 
is as follows.
If in a compact dimension with a periodic background vev of period $\omega$, we force the size of 
the extra dimension to be also $\omega$, all the fermionic states of period $2\omega$
will be automatically dismissed by the $\omega$ periodicity condition.
This imposition now seems rather arbitrary and awkward, since it does not come from any symmetry, 
but is, as we now know, extremely constraining.
Of course, the size of the compact space must be a multiple of the background period $\omega$, and 
because $2\omega$ is the minimum period that contains all the possible KK modes of all matter fields,
it seems natural for $2\omega$ to be the size of the extra dimension.
Close to the limit ($k\to1$), we can see how some of the ($k=1$) ``fat-brane'' modes split into period
 $\omega$ and period $2\omega$ modes.
The structure follows a curious pattern:
\bit
\item The zero-mode $\psi_0$ (of period $\omega$) and the first-level-mode $\psi_1$ (of period $2\omega$)
become degenerate (and massless) in the ($k\to1$) limit.
This means that when $k$ is close to 1, but still smaller than 1, we do have a chiral zero mode in the
 spectrum, but also an extremely light first-level massive fermion (as seen in Fig.~\ref{psi1models},
 when $k^2=0.99$ 
 (or $\omega=3.32 \omega_0$), we have $m_1^2\sim 10^{-49} m^2$). 
This feature seems to be also very advantageous for model building 
(e.g. explaining mass hierarchies or modeling the neutrino sector).
\item The next fermion level, $\psi_2$, of period $\omega$ 
(for $\delta>1$ and values of $k$ still close to 1), comes as 
a single state, with mass squared $\sim (\delta -1/2)m^2$.
\item If ($\delta>2$), there is at least another limiting discrete bound state. We see that, 
in this case, 
three fermion modes, $\psi_3$, $\psi_4$ and $\psi_5$, get close together and become nearly degenerate, 
when $k$ is close to 1, the mass splitting between them being extremely small compared to the gap between
them and the previous mode $\psi_2$. This feature again seems quite interesting.
\item
The next level, if $\delta$ is big enough, comes alone, and the next three together. This pattern 
repeats itself until the levels reach the continuum limit of the ($k=1$) case. Then all the modes 
become roughly evenly separated, and for the higher modes we recover the typical tower of masses 
following the $n^2\over 4$ structure\footnote{When the mass of the modes becomes large enough, we 
can understand this since the momentum along the extra dimension is so high that the modes do not 
feel the non-trivial background. They are therefore only limited by the periodicity of the 
compact space. The high-mass modes must therefore be more and more like sines and cosines, and their 
masses must follow the $n\over 2\omega$ spectrum.}.
\eit

\begin{figure}[h]
\hspace{-1.6cm}
\includegraphics[width=18.7cm]{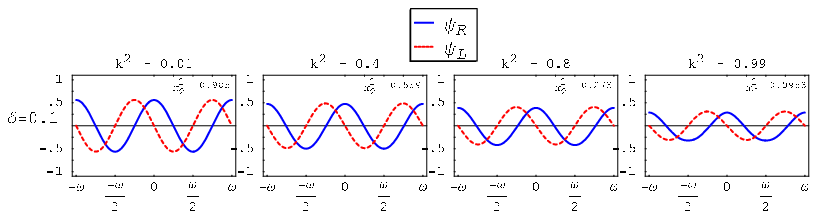}

\hspace{-1.8cm}
\includegraphics[width=19cm]{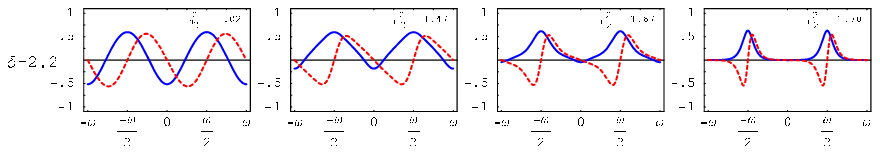}

\hspace{-1.8cm}
\includegraphics[width=19cm]{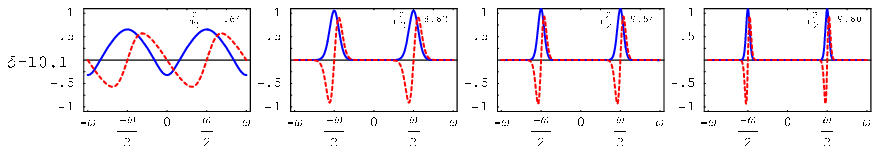}
\caption{The profile of the normalized mode $\psi^{\rm R/L}_2$, of period $\omega$, for $k^2=0.01,\ 0.4,\ 0.8$ 
and $0.99$ (or $\omega =1.01\omega_0,\ 1.34\omega_0 ,\ 1.93\omega_0$ and $3.32\omega_0$, with $\omega_0= 2 \pi / m$) 
and with $\delta=0.1,\ 2.2$ and $10.1$. The mass eigenvalues are given in units of $m^2$.  }
\label{psi2models}
\end{figure}

\begin{figure}[h]
\hspace{-1.6cm}
\includegraphics[width=18.7cm]{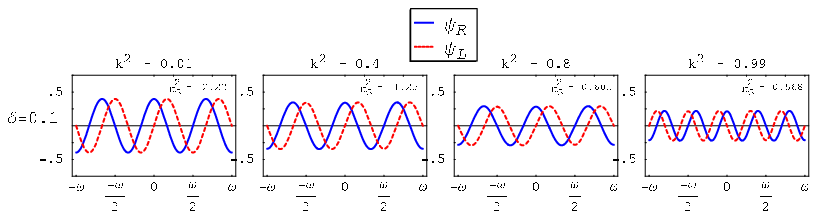}

\hspace{-1.8cm}
\includegraphics[width=19cm]{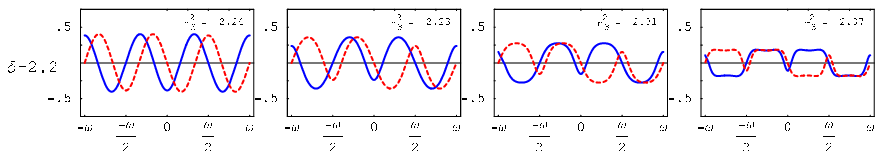}

\hspace{-1.8cm}
\includegraphics[width=19cm]{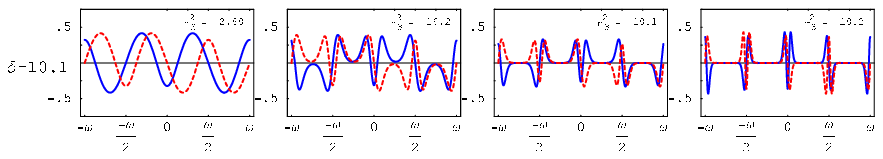}
\caption{The profile of the normalized mode $\psi^{\rm R/L}_3$, of period $2\omega$, for $k^2=0.01,\ 0.4,\ 0.8$ 
and $0.99$ (or $\omega =1.01\omega_0,\ 1.34\omega_0 ,\ 1.93\omega_0$ and $3.32\omega_0$, with $\omega_0= 2 \pi / m$) 
and with $\delta=0.1,\ 2.2$ and $10.1$. The mass eigenvalues are given in units of $m^2$. }
\label{psi3models}
\end{figure}

\begin{figure}[h]
\hspace{-1.6cm}
\includegraphics[width=18.7cm]{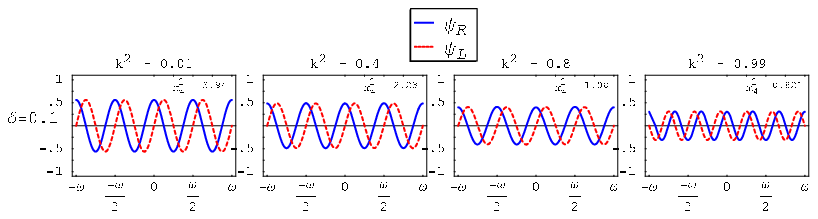}

\hspace{-1.8cm}
\includegraphics[width=19cm]{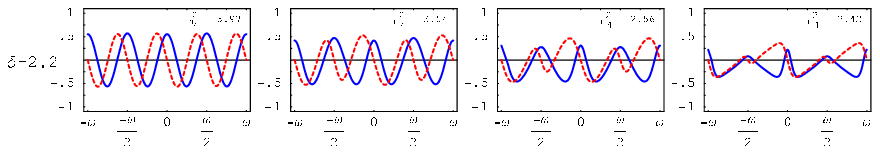}

\hspace{-1.8cm}
\includegraphics[width=19cm]{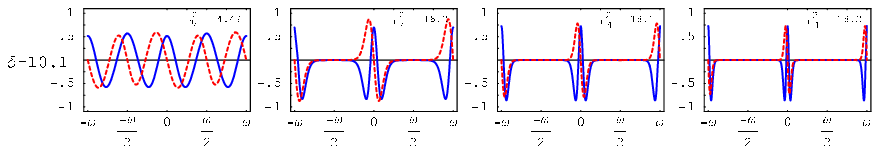}
\caption{The profile of the normalized mode $\psi^{\rm R/L}_4$, of period $\omega$, for $k^2=0.01,\ 0.4,\ 0.8$ 
and $0.99$ (or $\omega =1.01\omega_0,\ 1.34\omega_0 ,\ 1.93\omega_0$ and $3.32\omega_0$, with $\omega_0= 2 \pi / m$) 
and with $\delta=0.1,\ 2.2$ and $10.1$. The mass eigenvalues are given in units of $m^2$.  }
\label{psi4models}
\end{figure}

Figures \ref{psi1models}, \ref{psi2models}, \ref{psi3models} and \ref{psi4models} show the first 
modes $\psi_1^{\rm L/R}$, $\psi_2^{\rm L/R}$, $\psi_3^{\rm L/R}$ and $\psi_4^{\rm L/R}$ 
and their mass, for different values of $\delta$ and different compactification scales.

\section{Conclusions}
\label{com}

We have discussed the low-energy consequences of a non-trivial real scalar-field background in the context
of one compact extra dimension. The periodic background that appears within (1+4)-dimensional
$\phi^4$ theory was found, this solution being the analogue of the kink--antikink
approximate solution discussed so far in the literature. We have determined analytically the excitations
above the background and their spectrum. 
It was found that, in the presence of the non-trivial solution, there exists a
minimal size of the extra dimension that is determined by the mass parameter of the scalar potential:
$L_0=2\omega_0=4\pi/m$. We have shown
that imposing orbifold antisymmetry boundary conditions allows the elimination of a negative 
mass squared Kaluza--Klein ground-state mode that otherwise would cause an instability of the system.
The localization of fermionic modes in the presence of the non-trivial background was discussed in detail.
A simple and exact solution for the zero-mode fermionic states was found and the solution for
non-zero modes in terms of trigonometric series was constructed. The fermionic mass spectrum turned out
to be very interesting, with possible phenomenological consequences for constructing a realistic
theory based on a non-trivial background solution. 
Limiting cases of small harmonic 
oscillations and very large extra dimensions (the case considered in the literature so far)
were discussed and used as a test of the
general solutions for the fermionic Kaluza--Klein modes found in this work. 
We have shown that the natural size of the extra dimension
is twice as large as the period of the scalar background solution.

\vspace*{0.3cm}
\centerline{ACKNOWLEDGEMENTS}

\vspace*{0.3cm}

B.G. thanks Mikhail Shaposhnikov for an useful discussion on the periodic kink-like solutions.
The authors thank John Gunion for his contribution at the early stages of this work.
B.G. is supported in part by the State Committee for Scientific Research under grant 5~P03B~121~20 
(Poland). 
M.T. is supported by the Davis Institute for High Energy Physics. This work was also supported by a 
joint Warsaw--Davis
collaboration grant from the National Science Foundation and the Polish Academy of Science.

\renewcommand{\theequation}{A-\arabic{equation}}
\setcounter{equation}{0}  

\section*{Appendix}  

We are interested in finding all the periodic solutions of the equation:
\bea
\left[-\partial_y^2 + \left(f \phic\right)^2 - f \phic'- m^2_{n}\right] \psi^{\rm R}_y = 0\,.
\label{AppRy}
\eea
With these solutions, we can immediately get solutions to 
\bea
\left[-\partial_y^2 + \left(f \phic\right)^2 + f \phic'- m^2_{n}\right] \psi^{\rm L}_y = 0\,,
\label{AppLy}
\eea
as explained in the text, by simply doing a space translation of $\omega/2$.

It will prove useful to perform the redefinition 
\bea
\psi_y^{\rm R}(y)= v_y^{\rm R}(y) \exp\left\{- f \int^y_{y_0} d{\tilde{y}}\ \phic( \tilde{y})\right\} \,.
\label{psitransf}
\eea
We then have an equation for $v_y^{\rm R}(y)$:
\bea
-{d^2\over dy^2}  v_y^{\rm R} + 2 f \phic {d\over dy} v_y^{\rm R} -   m^2_{n} v_y^{\rm R} =0 \,.
\label{veq}
\eea
We see that it is easy to solve either Eq.~(\ref{veq}) or Eq.~(\ref{RL1}) for the zero-mode fermion ($m_n=0$). 
The chosen boundary conditions (\ref{fer_boud}) imply that the solution reads:
\beq
\psi_0^{\rm R}(y) \propto \exp\left\{- f \int_{y_0}^y d{\tilde{y}}\ \phic( \tilde{y})\right\}
\lsp {\rm and} \lsp
 \lsp \psi_0^{\rm L}(y)=0 \,.
\label{Appzeromode1}
\eeq
Now, using the background solution $\phic$ from Eq.~(\ref{periodicvev}) we get the unnormalized chiral mode
\bea
\psi_0^{\rm R}(y) \propto  \left[ {\rm dn}\!\left({m(y-y_0)\over \sqrt{k^2+1}} ,k^2\right) 
+ k\ {\rm cn}\!\left({m(y-y_0)\over \sqrt{k^2+1}},k^2\right) \right]^\delta \;\; {\rm and} \;\; 
\psi_0^{\rm L}(y) =0\,,
\label{ApppsiRzero}
\eea
where $\delta = \sqrt{2 f^2 / \lambda}$. This solution is exact.

If we now insert the periodic background solution $\phic$ in Eq.~(\ref{veq}) we get the 
Picard elliptic equation~\cite{Magnus} for the function $v_y^{\rm R}$:
\bea
-{d^2\over dy^2}  v_y^{\rm R} + 2 \delta m \sqrt{ k^2 \over k^2+1}\   {\rm sn}\left({m (y-y_0) \over 
\sqrt{k^2 +1}} ,\ k^2 \right) {d\over dy} v_y^{\rm R} -   m^2_{n} v_y^{\rm R} =0 \label{veq2}
\eea
Before going any further, we now need to use the known transformation formula for the elliptic sine function:
\bea
  {\rm sn}\left((1+\sqrt{1-\tilde{k}^2})\ u ,\ \left[{1-\sqrt{1-\tilde{k}^2}\over 1+\sqrt{1-\tilde{k}^2} }
\right]^2  \right) = (1+\sqrt{1-\tilde{k}^2})\ {{\rm sn}\left(u,\tilde{k}^2\right)\  {\rm cn}
\left(u,\tilde{k}^2\right)\over   {\rm dn}\left(u,\tilde{k}^2\right)} 
\eea
With this at hand we now redefine $u= {m\over 2} {(1+k)\over \sqrt{k^2+1}} (y-y_0)$ and we obtain:
\bea
-{d^2\over du^2}  v_u^{\rm R} + 2 \delta\  \tilde{k}^2     {{\rm sn}\left(u,\tilde{k}^2 \right) 
{\rm cn}\left(u,\tilde{k}^2\right) \over   {\rm dn}\left(u,\tilde{k}^2\right)}  {d\over du} v_u^{\rm R} -   
\beta_{n} v_u^{\rm R} =0 \label{vequ}
\eea
where $\tilde{k}^2 = {4 k \over (1+k)^2}$ and $\beta_{n} =4 {(k^2+1)\over (k+1)^2} {m_n^2\over m^2}$.

Now this can finally be brought to a simplified form of the generalized Ince equation\footnote{It 
could also take the form of the associated Lam\'e equation, by defining 
$w_u=v_u\  {\rm dn}\left(u,\tilde{k}^2\right)^{-\delta}$. This form is less interesting: in order to 
find general solutions, we have convert it back to the well studied generalized Ince equation 
and then expand in trigonometric series~\cite{Magnus}.}\cite{Magnus} by a new change of variables, 
$\ \sin{t} =  {\rm sn}\left(u,\tilde{k}^2 \right)$:
\bea
-(1-\tilde{k}^2 \sin^2{t}) {d^2\over dt^2}  v_t^{\rm R} + (1 + 2\delta)\   \tilde{k}^2  \sin{t} \cos{t}\  
{d\over dt} v_t^{\rm R}- \beta_{n} v_t^{\rm R} &=& 0 \,. 
\label{veqt}
\eea
Written in its canonical form we have
\bea
( 1 + a \cos{2t}) {d^2\over dt^2}  v_t^{\rm R} + b \sin{2t} {d\over dt} v_t^{\rm R} - \lambda_{n} 
v_t^{\rm R} &=& 0 \,,
\label{Ince}
\eea 
where $\ a={\tilde{k}^2\over 2-\tilde{k}^2}={2 k\over k^2+1}\ ,\ b=-(1 + 2\delta){2 k\over k^2+1}\  $ 
and $\lambda_n={(k+1)^2\over k^2+1}\beta_n=4 {m_n^2\over m^2} $.

We will now recall some known results related to the special case of the generalized Ince equation 
from \cite{Magnus}.
It is known that, for any real values of $a$ (satisfying $|a|<1$) and $b$, there exist infinitely many values 
of the parameter $\lambda_n$ such that Eq.~(\ref{Ince}) has an even or odd periodic solution, of 
period $\pi$ or $2\pi$. We will denote them as $\lambda^\pi_{2n}$, $\lambda^{2\pi}_{2n+1}$ and 
$\lambda^{2\pi}_{2n+2}$, where $n=0,1,2,3...$
In our case, for each of the $\lambda_{2n}^\pi$, there must be two linearly independent such solutions 
(except for the ground state $n=0$, which allows only one periodic solution).
If and only if the parameter $\delta$ is an integer, there will be also two linearly independent periodic 
solutions of period $2\pi$, and then $\lambda^{2\pi}_{2n+1}=\lambda^{2\pi}_{2n+2}$. When $\delta$ is not 
an integer, then one of the two linearly independent solutions will be periodic, of period $2\pi$, but the 
other one will not be periodic.
The $\lambda_n$'s satisfy the inequalities:
\bea
\lambda^\pi_0 < \lambda^{2\pi}_1 \leq  \lambda^{2\pi}_2 < \lambda^\pi_2 <  \lambda^{2\pi}_3 \leq  
\lambda^{2\pi}_4 < \lambda^\pi_4 < \cdots \,,
\eea
where the equalities hold whenever $\delta$ is integer. In our specific case we have $\lambda^\pi_0 = m_0 =0$.

Since the solutions to Eq. (\ref{Ince}) are even or odd and of period $\pi$ or $2\pi$, the solutions can 
be given with trigonometric series:
\bea
CI^{2n}_\pi(t)=\sum^{\infty}_{r=0} A^{2n}_{2r} \cos{2r t}\hspace{1.5cm}   &\lsp&\  SI_\pi^{2n}(t)=
\sum^{\infty}_{r=1} B^{2n}_{2r} \sin{2r t}\label{Ipi}  \\
CI^{2n+1}_{2\pi}(t)=\sum^{\infty}_{r=0}C^{2n+1}_{2r+1} \cos{(2r+1)t} &\lsp& SI^{2n+2}_{2\pi}(t)=
\sum^{\infty}_{r=0} D^{2n+2}_{2r+1} \sin{(2r+1) t}\ \ \ \ \label{I2pi} 
\eea
where $n=0,1,2,3,\dots$ The $CI$'s are even while the $SI$'s are odd functions of $t$. The periods are as follows:
$\pi$ for $CI^{2n}_\pi$, $SI^{2n}_\pi$ and $2\pi$ for $CI^{2n+1}_{2\pi}$, $SI^{2n+2}_{2\pi}$ (they correspond to
the periods $\omega$ and $2\omega$, respectively, in terms of the variable $y$).
Now, we define
\bea
Q(r)&=&2a r^2 - br\ =\ {4k\over k^2+1}\  r\ \left(r + \delta + 1/2 \right)\\
Q^*(r)&=&2 Q(r-1/2) =\ {4k\over k^2+1}\ (2r-1)\ (r + \delta) \,.
\eea
The parameters $ A^{2n}_{2r}$, $B^{2n}_{2r}$, $C^{2n+1}_{2r+1}$ and $D^{2n+2}_{2r+1}$ satisfy the 
following three term recurrence relations \cite{Magnus}:
\bea
-\lambda^\pi_{2n} A_{0} + Q(-1) A_{2}&=&0\\
Q(r-1)  A_{2r-2} + \left(4r^2-\lambda^\pi_{2n}\right) A_{2r} + Q(-r-1) A_{2r+2}&=&0 \hspace{3cm}\label{As}\\
r=1,2,3, ...&&\non
\eea
\bea
\left(4-\lambda^\pi_{2n}\right) B_{2} + Q(-2) B_{4}&=&0\\
Q(r-1)  B_{2r-2} + \left(4r^2-\lambda^\pi_{2n}\right) B_{2r} + Q(-r-1) B_{2r+2}&=&0 \hspace{3cm}\label{Bs}\\
r=2,3, ...&&\non
\eea
\bea
\left[Q^*(0) - 2(\lambda^{2\pi}_{2n+1}-1)\right] C_{1} + Q^*(-1) C_{3}&=&0\\
Q^*(r)  C_{2r-1} + 2 \left[(2r+1)^2-\lambda^{2\pi}_{2n+1}\right] C_{2r+1} + Q^*(-r-1) C_{2r+3}&=&0 
\hspace{3cm}\label{Cs}\\
r=1,2,3, ...&&\non
\eea
\bea
\left[-Q^*(0) - 2(\lambda^{2\pi}_{2n+2}-1)\right] D_{1} + Q^*(-1) D_{3}&=&0\\
Q^*(r)  D_{2r-1} + 2\left[(2r+1)^2-\lambda^{2\pi}_{2n+2}\right] D_{2r+1} + Q^*(-r-1) D_{2r+3}&=&0 
\hspace{3cm}\label{Ds}\\
r=1,2,3, ...&&\non
\eea
The condition that the series (\ref{Ipi}) and (\ref{I2pi}) converge will set the characteristic values 
of $\lambda^{\pi/2\pi}_i$. These characteristic equations can be written in terms of infinite 
determinants \cite{wang} or, as we will do, as infinite continued fractions \cite{Arscott}.

We need to remember that we are solving the equation for the right-hand fermion modes $\psi_{\rm R}^n$. 
The boundary conditions impose that these are even functions of $y$. So, we might be inclined to 
immediately disregard the solutions $SI_\pi^{2n}$ and $SI^{2n+2}_{2\pi}$.
However, it turns out that they are needed to obtain the left-hand 
modes $\psi_{\rm L}^n$, which are odd functions of $y$. We do this by translating the general solutions 
of Eq.~\ref{AppRy}, by $\omega/2$.

If we perform 
the translation of the period $\omega$ even solution of (\ref{AppRy}), constructed with $CI^{2n}_\pi$, 
we do get an even solution of (\ref{AppLy}). But we want $\psi_{\rm L}$ to be odd, and that solution is 
thus not good. What is needed is to compute the period $\omega$ odd solution of (\ref{AppRy}), constructed with 
$SI_\pi^{2n}$. The translated function will now be odd. As we said earlier, even and odd solutions 
of period $\pi$ (period $\omega$ in $y$ coordinate) of Eq.~(\ref{AppRy}) are guaranteed to exist for the same 
specific eigenvalues. 

The story is different for the period $2\omega$ solutions. Now the translation of the period $2\omega$ 
even solution of 
(\ref{AppRy}) constructed with $CI^{2n+1}_{2\pi}$ will give an odd solution of (\ref{AppLy}), which is 
exactly the solution for $\psi_{\rm L}$ that we need. Therefore 
the odd series $SI^{2n+1}_{2\pi}$ with coefficients $D_{2r+1}^{2n+2}$ will not be 
needed, since they correspond to odd solutions of (\ref{AppRy}) and to even solutions of (\ref{AppLy}), 
precisely the opposite parities needed.

Summarizing, the final solutions of (\ref{AppRy}) and (\ref{AppLy}) are as follows:
\bea
&\psi_{2n}^{\rm R}(y)   & \propto\ \psi_0(y)\ \ CI^{2n}_\omega\!\!\left[\ {\rm am}\!\left({m\over 2} {(1+k)\over 
\sqrt{k^2+1}}\ y,\ {4 k \over (1+k)^2} \right)\right]\\
&\psi_{2n+1}^{\rm R}(y) & \propto\ \psi_0(y)\ \ CI^{2n+1}_{2\omega}\!\!\left[\ {\rm am}\!\left({m\over 2} {(1+k)\over 
\sqrt{k^2+1}}\ y,\ {4 k \over (1+k)^2} \right)\right]  \\
&\psi_{2n}^{\rm L}(y)   & \propto\ {\psi_0(y+\omega/2)}\ \ SI^{2n}_\omega\!\!\left[\ {\rm am}\!\left({m\over 2} {(1+k)\over 
\sqrt{k^2+1}} (y+\omega/2),\ {4 k \over (1+k)^2} \right)\right] \\
&\psi_{2n+1}^{\rm L}(y) & \propto\ {\psi_0(y+\omega/2)}\ \  CI^{2n+1}_{2\omega}\!\!\left[\ {\rm am}\!\left({m\over 2} 
{(1+k)\over \sqrt{k^2+1}} (y+\omega/2),\ {4 k \over (1+k)^2} \right)\right]\ \ \ \ \ \ 
\eea
where
\beq
\psi_0(y)=\left[ {\rm dn}\!\left({m \over \sqrt{k^2+1}}\ y ,\ k^2\right) 
+ k\ {\rm cn}\!\left({m \over \sqrt{k^2+1}}\ y ,\ k^2\right) \right]^\delta\
\eeq
and $\omega = {4\sqrt{1+k^2}\over m} K(k^2) $ is the period of $\phic(y)$. Finally, ${\rm am}\!\left(x,m\right)$ 
is the Jacobian elliptic amplitude. 

In order to find specific values of the eigenvalues such that these periodic solutions exist,
we will only need two transcendental equations (infinite continued fractions equations) to find the values 
$\lambda^\pi_{2n}$ and $\lambda^{2\pi}_{2n+1}$.
To simplify our notation, we will introduce the continued fraction notation:
\bea
{a\over b + {c\over d+{e\over f+\cdots }}} = {a \over b + }\ {c\over d+ }\ {e\over f+ }\cdots
\eea
Using the recurrence relations (\ref{Bs}) and (\ref{Cs}) (see \S 3.6 in \cite{Arscott} for the procedure) 
we find the two transcendental equations that fix all the values of $\lambda^\pi_{2n}$ and 
$\lambda^{2\pi}_{2n+1}$:
\bea
{4 - \lambda_{2n}^\pi} = {Q(-2)Q(1)\over 16 - \lambda_{2n}^\pi - }\ {Q(-3) Q(2) \over  36 - 
\lambda_{2n}^\pi -}\ {Q(-4) Q(3) \over  64 - \lambda_{2n}^\pi -}\ \cdots \ {Q(-N) Q(N-1) \over  4N^2 - 
\lambda_{2n}^\pi -}\cdots
\eea
\bea
Q^*(0)+2- 2\lambda^{2\pi}_{2n+1} = { Q^*(-1) Q^*(1)\over 9 - \lambda^{2\pi}_{2n+1} -}\ {Q^*(-2) Q^*(2) 
\over 25 -  \lambda^{2\pi}_{2n+1} - }\ \cdots \ {Q^*(-N) Q^*(N) \over  (2N+1)^2 - \lambda_{2n+1}^{2\pi} -  }\cdots
\eea
These can easily be solved numerically.

\end{document}